\documentstyle[12pt]{article}
\begin{document}

\title {THE THOMAS PRECESSION AND THE TRANSFORMATION TO ROTATING FRAMES}
\author{Luis Herrera\thanks{Postal address: Apartado 80793, Caracas 1080A,
Venezuela; E-mail address: laherrera@telcel.net.ve}
 and Alicia Di Prisco\\
Escuela de F\'{\i}sica. Facultad de Ciencias.\\ Universidad Central de Venezuela. Caracas, Venezuela.\\
}
\date{}
\maketitle

\begin{abstract}
The Thomas precession is calculated using three different 
transformations to the rotating frame.
It is shown that for sufficiently large values of $v/c$, 
important differences in the predicted angle of precession appear, 
depending on the transformation used. For smaller values of $v/c$ 
these differences might be measured by extending the time of observation.
\end{abstract}

\newpage
\section{Introduction}
The Thomas preceession refers to the precession of a gyroscope along an 
arbitrary path in Minkowski space. It was originally calculated by Thomas 
\cite{Th}, when discussing the orbital motion of a spinning electron 
in an atom. We shall consider here the particular case of a circular orbit.

After the seminal work of Thomas, his result has been reobtained using many different approaches and techniques.
 However it is important to keep in mind, that any calculation of the Thomas precession  involves explicitly (or implicitly)
 not only the transformation law to the rotating frame , but what is more fundamental, a relationship between
the tangential velocity and the radial distance from the axis of rotation (see eq.(\ref{v}) below)), which is implied by the transformation law. In other words it is the very definition
of a rotating observer (and also that of ``rotation'' itself) which is at stake here. Different definitions lead to different results for Thomas precession, independently 
of the approach used to perform such calculations. 

Let us recall that since the early days of special relativity, the well known set of equations
\begin{equation}
t'=t \qquad;\qquad \rho'=\rho \qquad;\qquad 
\phi'=\phi-\omega \,t \qquad;\qquad z'=z 
\label{gal}
\end{equation}
has been adopted, relating the non-rotating coordinate system of the frame $S$ 
(in cylindrical coordinates) to the coordinates $t', \rho', \phi', z'$ of the 
frame $S'$ rotating uniformly about $z$-axis on the $\rho, \phi$ plane.

It should be emphazised that there is nothing ``wrong'' with (\ref{gal}) (hereafter referred to as GAL) as a coordinate transformation.
Obviously, this or any other regular transformation is acceptable, and any physical experiment whose result is expressed through invariants, should yield the same answer
for any observer. However, when we assume that GAL defines a rotating frame, we are adopting  a precise definition of a rotating observer. 
Different definitions of rotating observers are associated with different transformations. In other words, one thing is using GAL (which always can be done) and another,
 quite different thing is using GAL and assume that $S'$ defines a uniformly rotating frame  about $z$-axis on the $\rho, \phi$ plane. Therefore when we say that something 
(say a gyroscope) is rotating, we must specify what we mean by this. This may be done by specifying the relationship between the rotating and the inertial (non-rotating) observer 
(e.g. using GAL).

However the ``Galilean'' character of GAL (understood as a transformation to a rotating frame)
renders it questionable at the ultrarelativistic regime. Particularly 
troublesome is the fact that (\ref{gal}) implies that the Galilean 
composite law of velocities applies  in flagrant 
contradiction with special relativity \cite{Tr}--\cite{KiKr}.

In order to overcome this problem a different transformation law 
was independently proposed by Trocheris \cite{Tr} and Takeno \cite{Ta}.

In cylindrical coordinates the Trocheris-Takeno (TT) transformation reads
$$
\rho' = \rho \qquad ; \qquad z' = z
\label{roztt}
$$
\begin{equation}
\phi' = \phi \, \left(\cosh{\lambda}\right) - 
t \, \left(\frac{c}{\rho}\, \sinh{\lambda}\right)
\label{tt}
\end{equation}
$$
t' = t \, \left(\cosh{\lambda}\right) - \phi \, 
\left(\frac{\rho}{c} \, \sinh{\lambda}\right)
$$
with
$$
\lambda \equiv \frac{\rho \omega}{c}
$$
where primes correspond to the rotating frame and $c$ and $\omega$ denote the velocity of light and the  the angular velocity of the rotating frame defined by GAL, respectively .

It can be easily seen \cite{Tr},\cite{Ta}, 
that (\ref{tt}) leads to the relativistic law
of composition of velocities and yields for the velocity of a fixed 
point in $S'$, the expression
\begin{equation}
v=c\tanh{\lambda}
\label{v}
\end{equation}
sending to infinity the ``light cylinder''.

However, as it has been recently brought out \cite{He}, TT is 
not without its problems. In particular, the fact that proper 
time intervals are independent of $\rho$ ($g_{t't'}=c^2$) and 
that the length of a circle of radius $\rho$ measured in $S'$ is 
$2\pi\rho$, contradict ``relativistic'' intuition (see the 
letter of Einstein to J.Petzoldt, quoted in \cite{St}).

Accordingly a modification of TT was proposed in \cite{He} 
to remove those objections. The modified TT transformation (MTT) reads
\begin{equation}
t = \left(e^{-\lambda^2/2} \cosh{\lambda}\right) t' + 
\left(e^{\lambda^2/2} \frac{\rho'}{c} \sinh{\lambda}\right) \phi'
\label{MTTt}
\end{equation}
\begin{equation}
\phi = \left(e^{-\lambda^2/2} \frac{c}{\rho'} \sinh{\lambda}\right)  t'+ 
\left(e^{\lambda^2/2} \cosh{\lambda}\right) \phi' 
\label{MTTf}
\end{equation}
\begin{equation}
z=z' \, \qquad;\qquad \, \rho=\rho'
\label{zro}
\end{equation}
or, solving for $t'$ and $\phi'$
\begin{equation}
t' = e^{\lambda^2/2} \left[\left(\cosh{\lambda}\right)  t - 
 \left(\frac{\rho}{c} \sinh{\lambda}\right) \phi \right]
\label{MTTtp}
\end{equation}
\begin{equation}
\phi' = e^{-\lambda^2/2} \left[ \left(\cosh{\lambda}\right) \phi -
\left(\frac{c}{\rho} \sinh{\lambda} \right) t\right]
\label{MTTfp}
\end{equation}

Two points deserve to be stressed:
\begin{enumerate}
\item The expressions 
for unprimed (primed) coordinates do not follow from the expressions 
for primed (unprimed) coordinates, by changing 
$\omega \rightarrow -\omega$ 
and $\phi'^{\longrightarrow}_{\longleftarrow} \phi$ ; 
$t'^{\longrightarrow}_{\longleftarrow} t$.
This is a consequence of the fact that in the derivation of MTT, as given in \cite{He}, the group properties of the rotation
 have been broken, by admitting a linear term in $\omega$ in the expressions for $\frac{d\phi'}{d\omega}$ and  
$\frac{dt'}{d\omega}$ (see equation (20) in \cite{He}).
However, this is a desirable feature of these transformations   since, unlike the case of Lorentz transformations 
between two inertial frames, now $S$ and $S'$ are physically different (one is 
inertial, whereas the other is not).
\item It should be clear that Minkowski spacetime remains the framework of the theory since all Riemann tensor
 components are vanishing. 
\end{enumerate}

To summarize: GAL and TT observers unlike MTT, define the inverse  transformation in the usual canonical way. On the other hand  MTT and TT, unlike GAL,
 send the ``light cylinder'' to infinity. Also,
unlike TT, both MTT and GAL  observers measure greater circles  as compared with the non-rotating observer.

It is the purpose of this work to bring out the implications that, different transformations to rotating frame, have
in the calculation of Thomas precession. Once again we must emphasize that the issue is not restricted to a simple coordinate transformation,
for in this case any invariant approach would give the same result. As mentioned before it is the very definition of
 rotation, associated with each transformation, which makes the difference in the predicted precession. 
In other words, using either GAL, TT or MTT, amounts to perform different experiments, since each one of these transformations implies a different definition of rotating observer.
Accordingly we shall now proceed to calculate the Thomas precession, 
providing at each step the preliminary results corresponding 
to each one of the transformations to be used (GAL, TT, MTT).

However, before closing this section, and for the benefit of the reader we list below the relevant components of the metric tensor in each system
(omitting primes).
\begin{eqnarray}
g^{MTT}_{tt}& = & c^2 e^{-\lambda^2}\nonumber \\
g^{MTT}_{\phi\phi} & = & -{\rho}^2 e^{\lambda^2}\nonumber \\
g^{MTT}_{t\phi} & = & 0
\end{eqnarray}

\begin{eqnarray}
g^{TT}_{tt}& = & c^2 \nonumber \\
g^{TT}_{\phi\phi} & = & -{\rho}^2 \nonumber \\
g^{TT}_{t\phi} & = & 0
\end{eqnarray}

\begin{eqnarray}
g^{GAL}_{tt}& = & c^2 (1-\lambda^2)\nonumber \\
g^{GAL}_{\phi\phi} & = & -{\rho}^2 \nonumber \\
g^{GAL}_{t\phi} & = & -2 \omega r^2
\end{eqnarray}

\section{The Thomas precession}
Let us start by defining the vorticity vector, which as usual 
is given by
\begin{equation}
\omega^\alpha = \frac{c}{2 \sqrt{-g}} \, \epsilon^{\alpha\beta\gamma\delta} 
u_\beta \omega_{\gamma\delta} = 
\frac{c}{2 \sqrt{-g}} \, \epsilon^{\alpha\beta\gamma\delta} 
u_{\beta} u_{\delta,\gamma}
\label{vv}
\end{equation}
where the vorticity tensor is given by
\begin{equation}
\omega_{\alpha\beta} = u_{\left[\alpha;\beta\right]} - \dot{u}_{[\alpha}u_{\beta]}
\label{vt}
\end{equation}
and $u_{\beta}$ denotes the four-velocity vector.

Next, for both MTT and TT the velocity vector of a point at rest in $S'$ 
(rotating with respect to $S$) is given by (in cilyndrical coordinates 
$x^{0,1,2,3} = ct,\rho,\phi,z$)
\begin{equation}
u^\alpha = \left(\cosh{\lambda}, 0, \frac{\sinh{\lambda}}{\rho}, 0\right)
\label{utta}
\end{equation}
or
\begin{equation}
u_\alpha = \left(\cosh{\lambda}, 0, - \rho \sinh{\lambda}, 0\right)
\label{uttab}
\end{equation}
whereas for GAL, the corresponding expressions are
\begin{equation}
u^\alpha = \left(\frac{1}{\sqrt{1-\frac{\omega^2\rho^2}{c^2}}}, 0, 
\frac{\omega}{c \sqrt{1-\frac{\omega^2\rho^2}{c^2}}}, 0\right)
\label{utta}
\end{equation}
\begin{equation}
u_\alpha = \left(\frac{1}{\sqrt{1-\frac{\omega^2\rho^2}{c^2}}}, 0, 
- \, \frac{\omega \rho^2}{c \sqrt{1-\frac{\omega^2\rho^2}{c^2}}}, 0\right)
\label{uttab}
\end{equation}
then a simple calculation yields
\begin{equation}
\omega^\alpha = \left(0, 0, 0, \frac{c}{2\rho}(\cosh{\lambda}\sinh{\lambda} 
+ \lambda)\right)
\label{omtta}
\end{equation}
or
\begin{equation}
\omega_\alpha = \left(0, 0, 0, - \, \frac{c}{2\rho}(\cosh{\lambda}\sinh{\lambda}
 + \lambda)\right)
\label{omttb}
\end{equation}
for MTT and TT, and 
\begin{equation}
\omega^\alpha = \left(0, 0, 0, 
\frac{\omega}{1-\frac{\omega^2 \rho^2}{c^2}}\right)
\label{omga}
\end{equation}
\begin{equation}
\omega_\alpha = \left(0, 0, 0, - \, 
\frac{\omega}{1-\frac{\omega^2 \rho^2}{c^2}}\right)
\label{omgb}
\end{equation}
for GAL.

From(\ref{omtta})--(\ref{omgb}) the absolute value of $\omega^\alpha$ 
is obtained at once.
For MTT and TT, we get
\begin{equation}
\Omega\equiv\left(-\omega_\alpha \omega^\alpha\right)^{1/2} = 
\frac{c}{2\rho} \left(\sinh{\lambda}\cosh{\lambda} + \lambda\right)
\label{Ott}
\end{equation}
whereas for GAL we obtain
\begin{equation}
\Omega = \frac{\omega}{1 - \frac{\omega^2 \rho^2}{c^2}}
\label{Og}
\end{equation}
which is a known result \cite{RiPe}.

Next, since $\Omega$ measures the rate of rotation with respect 
to proper time of world lines of points at rest in $S'$, 
relative to the local compass of inertia, then $-\Omega$ describes 
the rotation of the compass of inertia (the ``gyroscope'') with 
respect to reference particles at rest in $S'$ (see \cite{RiPe} 
for detailed discussion on this point). Therefore, after one 
complete revolution the change of orientation of the gyroscope 
as seen by the observer in $S'$, is given by
\begin{equation}
\Delta \phi' = - \Omega \Delta \tau'
\label{or}
\end{equation}
where $\Delta\tau'$ is the proper time interval (in $S'$) 
corresponding to the period of one revolution.

Now, for both MTT and TT the interval of coordinate time $t$ 
corresponding to the period of one revolution is
\begin{equation}
\Delta t=\frac{2 \pi \rho}{v}=\frac{2 \pi \rho}{\xi \rho}=\frac{2 \pi}{\xi}
\label{dt}
\end{equation}
where the ``angular velocity'' $\xi$ in now given by 
(see \cite{Tr}, \cite{Ta}, \cite{He}).
\begin{equation}
\xi=\frac{d\phi}{dt}=\frac{c}{\rho} \tanh{\lambda}
\label{xi}
\end{equation}
coinciding with $\omega$ for sufficiently small $\lambda$.

Obviously for GAL the corresponding expression is
\begin{equation}
\Delta t=\frac{2\pi}{\omega}
\label{dtg}
\end{equation}

Next, from(\ref{tt}) and (\ref{MTTt})--(\ref{MTTfp})
we obtain
\begin{equation}
\Delta t' = \frac{\Delta t}{\cosh{\lambda}}
\label{dttt}
\end{equation}
and
\begin{equation}
\Delta t' = \frac{\Delta t \,\, e^{\lambda^2/2}}{\cosh{\lambda}}
\label{dtmtt}
\end{equation}
for TT and MTT respectively, whereas
\begin{equation}
\Delta t' = \Delta t 
\label{dtga}
\end{equation}
for GAL.

Now, the relation between coordinate time interval and 
proper time interval (in $S'$), for TT is
\begin{equation}
\Delta \tau' = \Delta t' 
\label{pttt}
\end{equation}
whereas for MTT and GAL we have respectively
\begin{equation}
\Delta \tau' = e^{-\lambda^2/2} \; \Delta t' 
\label{ptmtt}
\end{equation}
and 
\begin{equation}
\Delta \tau' = \sqrt{1-\frac{\omega^2 \rho^2}{c^2}} \; \Delta t' 
\label{ptg}
\end{equation}
From the expressions above it follows that
\begin{equation}
\Delta\phi' = - \frac{2 \pi \Omega}{\xi \cosh{\lambda}}
\label{dft}
\end{equation}
for both TT and MTT, and 
\begin{equation}
\Delta\phi' = - \frac{2 \pi}{\left(1-\frac{\omega^2 \rho^2}{c^2}\right)^{1/2}}
\label{dfg}
\end{equation}
for GAL

In this later case (GAL), since the rotating system 
precesses the baseline by $2\pi$ per revolution, then the precession 
per revolution relative to $S$ is given by
\begin{equation}
\Delta \phi = \Delta \phi' + 2\pi = 
2\pi\left[1 - \frac{1}{\left(1-\frac{\omega^2 \rho^2}{c^2}\right)^{1/2}}\right]
\label{pre}
\end{equation}
which is the well known Thomas result \cite{RiPe}.

The calculation for TT and MTT requires some care. Thus, let us assume that at
\begin{equation}
t'_1=t'_0 \qquad;\qquad
z'_1=z'_0 \qquad;\qquad
\rho'_1=\rho'_0
\label{1}
\end{equation}
the orientation of the gyroscope is given by
\begin{equation}
\phi'_1=\phi'_0
\label{fi1}
\end{equation}
and, at
\begin{equation}
t'_2=t'_0 +\Delta t' \qquad;\qquad
z'_2=z'_0 \qquad;\qquad
\rho'_2=\rho'_0
\label{2}
\end{equation}
by
\begin{equation}
\phi'_2=\phi'_0 + \Delta \phi'
\label{fi2}
\end{equation}
where $\Delta t'$ and $\Delta \phi'$ are given by (\ref{dttt})(TT) 
(or (\ref{dtmtt})(MTT)) and (\ref{dft}).

Then using (\ref{tt}) and (\ref{MTTt})--(\ref{MTTfp}) we find for 
the total change of orientation of the gyroscope after one revolution 
as measured in $S$ 
\begin{equation}
\Delta \phi = 2 \pi \left[1-\frac{e^{\lambda^2/2}}{2 \tanh{\lambda}} 
\left(\sinh{\lambda} \cosh{\lambda} + \lambda\right)\right]
\label{Tmtt}
\end{equation}
for MTT, and
\begin{equation}
\Delta \phi = 2 \pi \left[1-\frac{1}{2 \tanh{\lambda}} 
\left(\sinh{\lambda} \cosh{\lambda} + \lambda\right)\right]
\label{Ttt}
\end{equation}
for TT.

Before closing this section, the following comments are in order:

Observe that $\Omega_{TT}=\Omega_{MTT}$ and furthermore, as it follows from (25)-(30), the amount of proper time corresponding to a time interval $\Delta t$ in $S$, is the same in both
 $TT$ and  $MTT$ frames. Accordingly the change of orientation of the gyroscope for a fixed amount of proper time as measured by each of the observers ($TT$ and $MTT$) is the
same, as it follows from (21). Specifically, up to order of $\lambda^2$, we have
\begin{equation}
\Omega_{TT,MTT}\approx\omega(1 + \lambda^2/3)
\label{c1}
\end{equation}

giving

\begin{equation}
\Delta\Phi_{MTT,TT}^{\prime}\approx-\omega \Delta t (1-\lambda^2/6).
\label{c2}
\end{equation}

Of course the total change of orientation  after one revolution in S, is different as calculated with $TT$ or $MTT$ as it follows from (38) and (39).

Now, for $GAL$ the situation is different. Indeed, even up to order $\lambda^2$, $\Omega_{GAL}$ differs from $\Omega_{TT,MTT}$, as we have
\begin{equation}
\Omega_{GAL}\approx\omega (1 + \lambda^2)
\label{c2}
\end{equation}

yielding

\begin{equation}
\Delta\Phi_{GAL}^{\prime}\approx-\omega \Delta t (1 + \lambda^2/2).
\label{c3}
\end{equation}

Where the fact has been used that the proper time (for $GAL$) corresponding to an  interval $\Delta t$ in $S$, is

\begin{equation}
\Delta\tau_{GAL}\approx\Delta t (1-\lambda^2/2).
\label{c4}
\end{equation}

which up to this order ($\lambda^2$) is the same as $\Delta\tau_{MTT,TT}$.
\section{Conclusions}

We have seen so far that different transformations to rotating frames, implying different definitions of rotating observer,  lead to different expressions for
the total precession of a gyroscope on a circular orbit.

Now, expanding (\ref{pre}), (\ref{Tmtt}) and (\ref{Ttt}) in power 
series of $\lambda$ up to quadratic terms, one obtains
\begin{equation}
\Delta \phi \approx - \pi \frac{v^2}{c^2}
\label{RGAL}
\end{equation}
\begin{equation}
\Delta \phi \approx - \frac{4 \pi}{3} \frac{v^2}{c^2}
\label{RTT}
\end{equation}
\begin{equation}
\Delta \phi \approx - \frac{7 \pi}{3} \frac{v^2}{c^2}
\label{RMTT}
\end{equation}
for GAL, TT and MTT respectively. Thus, for sufficiently 
small values of $\lambda$, the differences in the predicted 
value of $\Delta \phi$ after one period are of the order $\pi \frac{v^2}{c^2}$ 
(see figure 1). However since the effect is additive after each revolution, 
it may become important when observing the system for sufficiently long time.

For values of $\frac vc$ approaching the ultrarelativistic regime those differences 
are quite important after one period, as exhibited in figure 2.

We ignore about the technical feasibility to carry on an experiment allowing 
the detection of such differences. However, should this be possible, then 
a test for the correct transformation to rotating frame would be at hand.

We would like to stress the fact that this problem is not only of 
academic interest, in view of future navigation and time transfer systems 
which are now being contemplated with sub-nanosecond accuracy (see 
\cite{Ba} and references therein).

Finally, it is worth remembering that  Thomas precession plays a fundamental role in the theory of spin-orbit coupling, and consequently it is involved in all
 related effects, such as Zeeman effect. Unfortunately all calculations we have found in literature  describing this
effect (and others involving spin-orbit coupling) , not only are perturbative and to order $(v/c)^2$ but also include other approximations which render very difficult (to us at least) 
to contrast the differences expressed through our equations (\ref{RGAL})-(\ref{RMTT}) with the experimental data. However we do believe  that perhaps more competent people than we are, on this subject,
 could propose a gedanken (or a real) experiment involving spin-orbit coupling, and whose result could settle the question about the appropriate definition of a rotating observer.

\section{Figure captions}
Figure 1.
$\Delta \phi/2 \pi$ as function of $x(=v/c)$, calculated with 
three different tranformations (MTT,TT and Galilean).
\\
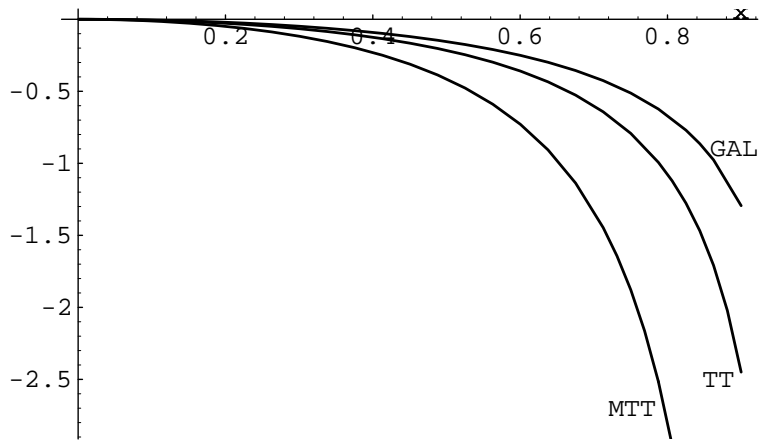
Figure 2.
Same as figure 1, but for larger values of $x$.

\section*{Acknowledgements}

The financial assistance from  M.C.T.,Spain, BFM2000-1322 is
gratefully acknowledged.

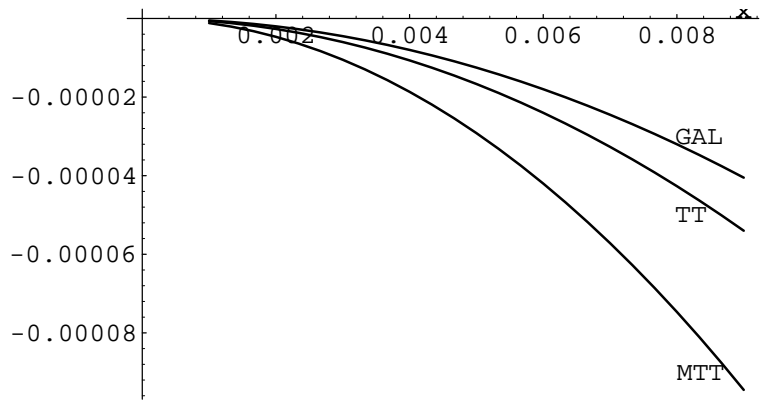
\end{document}